\documentclass[manuscript]{aastex}
\usepackage{color,graphicx,float}

\newcommand{\km}{\,\mathrm{km}}
\newcommand{\au}{\,\mathrm{au}}

\newcommand{\PSone}{\protect \hbox {Pan-STARRS~1}}
\newcommand{\meter}{\,\mathrm{m}}
\newcommand{\eg}{{\it e.g.}}
\newcommand{\TC}{{2008~TC$_3$}}
\newcommand{\ie}{{\it i.e.}}
\newcommand{\mags}{\,\mathrm{mag}}

\begin{document}

\title{The effect of parallax and cadence\\
on asteroid impact probabilities and warning times}

\author{
Peter Vere\v{s}\altaffilmark{1} (veres@ifa.hawaii.edu),
Davide Farnocchia\altaffilmark{2},
Robert Jedicke\altaffilmark{1},
Federica Spoto\altaffilmark{3},
}

\slugcomment{41 Pages, 11 Figures, N Table}

\altaffiltext{1}{Institute for Astronomy, 2680 Woodlawn Dr, Honolulu,
  HI, 96822}

\altaffiltext{2}{Jet Propulsion Laboratory, California Institute of
  Technology, 4800 Oak Grove Drive, Pasadena, CA, 91101}

\altaffiltext{3}{Department of Mathematics, University of Pisa, Largo
  Bruno Pontecorvo 5, 56127 Pisa, Italy}

\shorttitle{Effects of parallax and cadence on asteroid impact probabilities.}
\shortauthors{Vere\v{s} \etal}

\begin{abstract}

We study the time evolution of the impact probability for synthetic
but realistic impacting and close approaching asteroids detected in a
simulated all-sky survey.  We use the impact probability to calculate
the impact warning time ($t_{warn}$) as the time interval between when
an object reaches a Palermo Scale value of $-2$ and when it impacts
Earth.  A simple argument shows that $t_{warn} \propto D^x$ with the
exponent in the range $[1.0,1.5]$ and our derived value was $x=1.3 \pm
0.1$ .  The low-precision astrometry from the single simulated all-sky
survey could require many days or weeks to establish an imminent
impact for asteroids $>100\meter$ diameter that are discovered far
from Earth.  Most close approaching asteroids are quickly identified
as not being impactors but a size-dependent percentage, even for those
$>50\meter$ diameter, have a persistent impact probability of
$>10^{-6}$ on the day of closest approach. Thus, a single all-sky
survey can be of tremendous value in identifying Earth impacting and
close approaching asteroids in advance of their closest approach but
it can not solve the problem on its own: high-precision astrometry
from other optical or radar systems is necessary to rapidly establish
an object as an impactor or close approacher.  We show that the parallax afforded by surveying the sky from two sites is only of benefit for a small fraction of the smallest
objects detected within a couple days before impact: probably not
enough to justify the increased operating costs of a 2-site
survey. Finally, the survey cadence within a fixed time span is
relatively unimportant to the impact probability calculation.  We
tested three different reasonable cadences and found that one provided
$\sim10\times$ higher (better) value for the impact probability on the
discovery night for the smallest ($10\meter$ diameter) objects but the 
consequences on the overall impact probability calculation is negligible.

\end{abstract}
\keywords{Near-Earth Objects; Asteroids, Dynamics}
\maketitle

\section{Introduction}

The past quarter century has witnessed an exponential increase in the
number of known near-Earth objects\footnote{NEOs are asteroids or
  comets that have a perihelion distance $<1.3\au$} (NEO)
accompanied by a concomitant improvement in the ability to calculate
their impact probabilities with Earth.  The job of identifying the
largest and most hazardous NEOs, those larger than about $1\km$
diameter, has mostly been accomplished and asteroid surveys are now
focussing on the individually less hazardous but far more numerous
smaller asteroids.  The large asteroids can be detected at great
distances years to centuries in advance of their impact but the
smaller asteroids may only be detected on their final approach, if at
all, since about 40\% of them must approach from the direction of the
Sun in daylight sky.  This work quantifies how the impact probability
and warning time evolve in the impact apparition for the smaller
asteroids as a function of their size, time after discovery, and
observing cadence.  In particular, we examine whether the parallax
afforded by observations at nearly the same time from two independent
observatories provides leverage in improving the impact probability
calculation or increasing the impact warning time.

The Catalina Sky Survey \citep[CSS,\ ][]{Larson1998} and
\PSone\ system \citep[Pan-STARRS; \eg][]{Kaiser2002,Hodapp2004}
currently dominate the field of NEO discovery --- almost 90\% of all
NEOs and about 75\% of all potentially hazardous objects\footnote{PHOs
  are NEOs that have a minimum orbital intersection distance
  \citep[\eg][]{Gronchi2005} with Earth of $<0.05\au$ and absolute
  magnitude $H<22$} (PHO) were discovered by these two surveys in
calendar years 2012 and 2013.  The known population of NEOs larger
than $1\km$ diameter is $>90$\% complete \citep{Mainzer2011c} so the
discovery rate of NEOs in this size range has decreased by about a
factor of 6 from a peak of 93 in the year 2000 to about 15/year in the
last two years.

Despite the success of the surveys in the past few decades it remains
the case that the most likely warning time for an impact is {\it zero}
--- contemporary surveys are unlikely to detect smaller but still
dangerous asteroids because they do not survey the entire sky deeply
or regularly enough to identify the next impactor.  The surveys are
further limited by the simple fact that ground-based facilities can
not survey during the day and about 40\% of all impactors will
approach from the direction of the Sun.  These problems were
spectacularly highlighted by the Chelyabinsk impact on the morning of
15 February 2013 \citep[\eg][]{Brown2013,Borovicka2013} --- with
absolutely no warning a $\sim17\meter$ diameter object blew up in the
atmosphere with an energy equivalent to about 500\,kilotons of TNT,
damaging buildings 50\,km away in the city of Chelyabinsk and injuring
about 1,500 people.

The impact risk associated with the unknown objects $>1\km$ diameter
is now comparable to the impact risk with the much more numerous but
individually less destructive objects with diameters $<1\km$.  The new
balance in the impact risk, along with the realization that smaller
impacts may be more numerous but less destructive than anticipated a
decade ago \citep[\eg][]{Brown2013}, has contributed to an increased
interest and funding for the NEO survey programs in recent years.
\eg\ NASA's NEO Observations (NEOO) program office now
solicits\footnote{ROSES 2011 NEOO solicitation section C.9.1.1.}
proposals for surveys that `provide capability to detect the subset of
90\% of PHOs down to 140 meters in size'

The smaller NEOs are more difficult to detect than the larger ones,
will be detected ({\it if} they are detected) closer to Earth, and
consequently have shorter observational arcs.  The limited time range
of the set of detections can make it difficult to identify real
impactors even during the apparition in which the impact will take
place.  This was not the case for the few-meter diameter asteroids \TC
\citep[\eg][]{Jenniskens2009} and 2014~AA\footnote{Minor Planet
  Electronic Circular 2014-A02}, the only natural objects to be
discovered before striking Earth.   The very smallest objects
will be discovered so close to Earth that, if individual detections of
the object can be associated with one another as a `tracklet'
\citep{Denneau2013}, the non-linear motion of the detections on the
sky-plane due to topocentric parallax can provide enough leverage in
the orbit solution to predict an impact.

The observable characteristics of NEOs that will impact Earth can be
quite different from those of other NEOs
\citep[\eg][]{Farnocchia2012,Veres2009,Chesley2004}.  For instance,
their observable steady-state distribution on the sky-plane is a
function of their size and time before impact.  Decades before impact
they tend to be concentrated in `sweet spots' near the ecliptic and
within about $120\arcdeg$ of the Sun.  As the time until impact
decreases from weeks to days they spread out over most of the sky but
there are still concentrations in the direction looking towards and
away from the Sun.  An object on its `death plunge' must be moving
directly towards Earth in a geocentric reference frame so that about a
week before impact its apparent rate of motion may be small --- likely
mimicking the rate of motion of much more distant and totally harmless
asteroids and perhaps not triggering followup that would allow an
impact probability calculation.

The techniques employed for the impact probability calculation have
evolved dramatically over the past few decades with the realization
that asteroid impacts have shaped the Moon's surface and influenced
the evolution of life on Earth.  Indeed, it was only 34 years ago that
\citet{Alvarez1980} proposed that the KT extinction was the result of
an asteroid impact and, even though \citet{Opik1952} stated that
``Over a dozen meteor craters are at present known on the earth's
surface'', it was only in 1960 that \citet{Chao1960} found strong
physical evidence that Meteor Crater in Arizona, USA, was formed in an
impact event.

\citet{Opik1952}'s estimated collision rates using the `Theory of
Probabilities' for the entire NEO population were surprisingly good
given that only six NEOs were known at the time.  His collision
probability formulae formed the basis of much of the impact collision
work in the next decades \citep[\eg][]{Bottke1996,Kessler1981} but
were eventually supplanted by new numerical techniques
\citep{Milani2002}.  The two primary operational asteroid impact
warning systems, the Jet Propulsion Laboratory's Sentry system and the
NEODyS CLOMON2 system, calculate the collision probability by
generating synthetic `Virtual Asteroids' (VA) on orbits that are
consistent with the known set of observations and propagating all of
them into the future with a N-body integrator to search for impacts
\citep{Milani2005}.  These impact warning systems are based on a
geometric sampling technique for which the identification of the Virtual
Impactors (VI) is performed on the line-of-variation
\citep[LOV,\ ][]{Milani2005b} thus avoiding the poor efficiency inherent
to the Monte Carlo methods, especially when the collision probability
is small.

Impact predictions are extremely sensistive to the orbit accuracy
which depends on many factors but the primary drivers are the length
of the observational arc and the astrometric accuracy
\citep[\eg\ ][]{Desmars2013}.  The longer the arc and the better the
astrometry the more accurate the orbit.  The latter effect is best
illustrated by radar detection of asteroids that provide exquisite
range and range-rate information thereby dramatically improving the
impact probability accuracy and/or extending the time frame during
which the impact probability can be calculated \citep{Ostro2002}.

Impactors can be either direct or resonant \citep{Milani1999}.  Direct
impactors collide with Earth during their first known encounter and
must be discovered far away to have a large warning time.  The warning
time for small impactors can be significantly less than one orbital
period because they have to be close to Earth to be detected.  On the
other hand, resonant impactors experience intervening Earth encounters
before collision.  The intervening encounters are the main source of
non-linearity in the dynamics and usually prevent a conclusive
assessment of the impact threat but provide additional observational
opportunities to detect and constrain their orbits and the impact
threat.

In this work we focus on the evolution of the collision probability
with time for a single survey and concentrate on direct Earth
impactors that are detected in the apparition during which the impact
occurs.  The smaller the asteroid the more likely this scenario as the
likelihood that small asteroids will be detected in earlier
apparitions is 1) small and, even if they are detected, 2) it is
unlikely that they will be recoverable in future apparitions because
of the large uncertainties in their ephemeris based on the short
observational arcs in the discovery apparition.  Thus, we concentrate
on collision probability evolution with time for $300\meter$,
$100\meter$, $50\meter$ and $10\meter$ diameter impactors.

We also explore whether the collision probability calculation
benefits from simultaneous or nearly-simultaneous parallax
measurements from two observatories.  The heliocentric motion of the
impactor and Earth as well as the topocentric rotation of the observer
about the geocenter produce `parallax' between successive observations
of the same object even from the same site. For very close objects
that will impact within days of discovery there may be benefits from
the two-site scenario --- especially in rapidly identifying the object
as an impactor.

Finally, we measure the single-system impact warning time as a
function of impactor diameter.  We expect the warning time to be
longer for larger objects but the exact relationship between diameter
and warning time is not intuitively obvious.  The larger objects are
discovered at greater distances where their rate of motion is similar
to the much more distant main belt objects and the impact probability
will be much smaller.  If the impact probability is too small it may
not cross the threshold to flag the object as an imminent impactor.

\section{Method}

\subsection{Synthetic asteroid populations}

\begin{figure}[btp]
\centering

\plotone{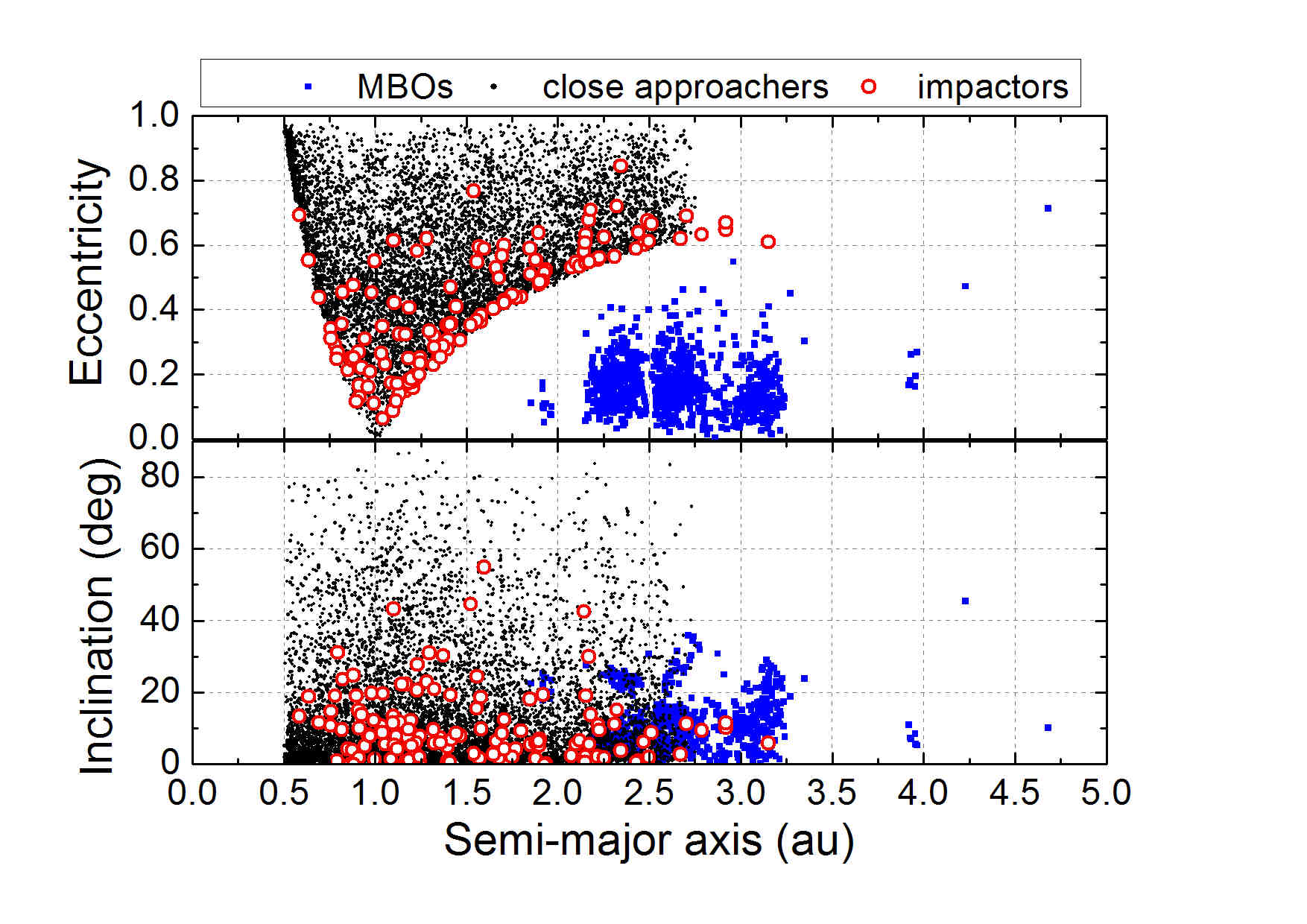}

\caption{Eccentricity (top) and inclination (bottom) vs. semi-major
  axis for synthetic impactors (circles), close approachers (black dots) and 
  main belt objects (grey squares).}
\label{fig.aei}
\end{figure}

\begin{figure}[btp]
\centering
\epsscale{1.0}
\plotone{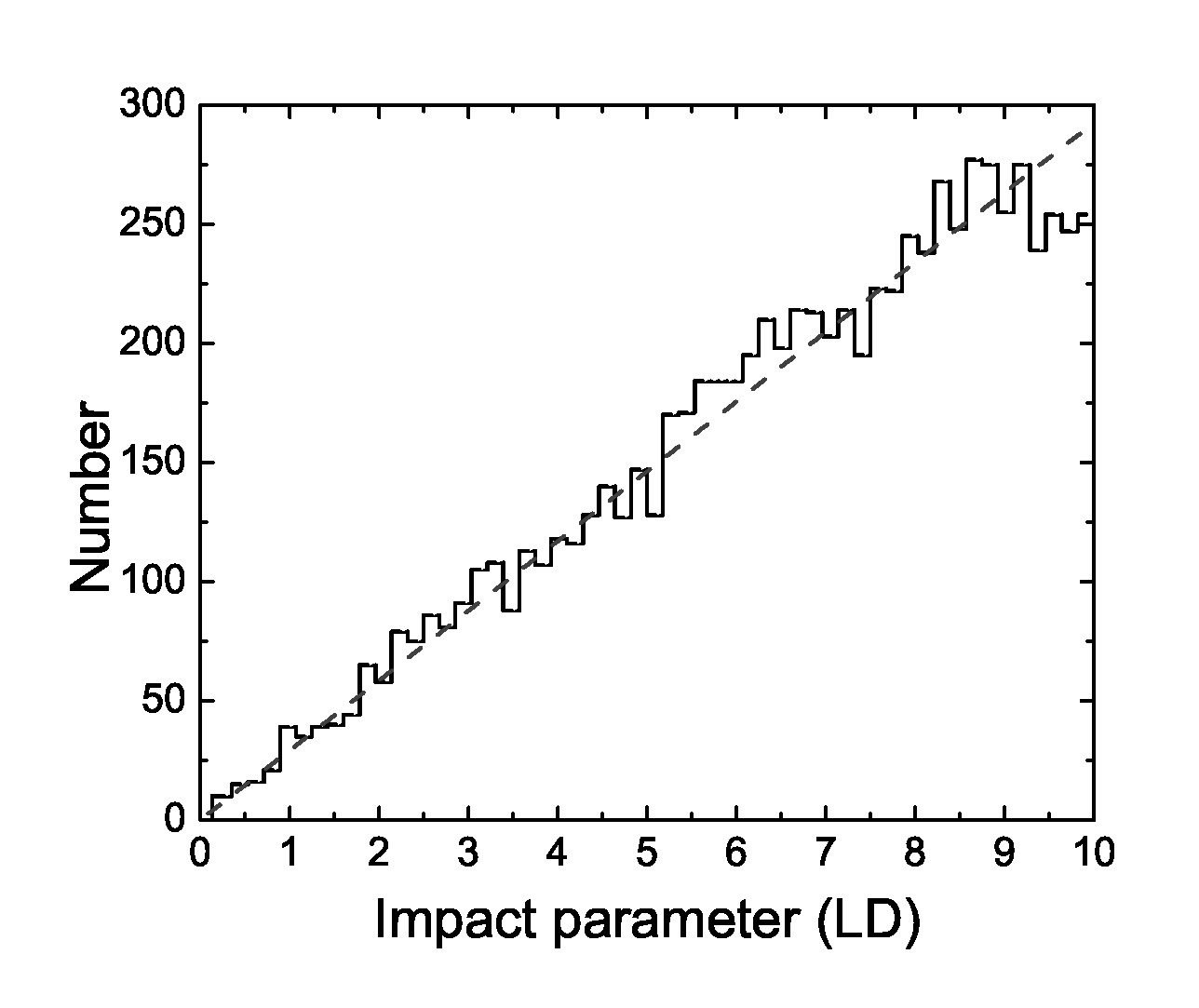}
\caption{Impact parameters (closest approach distance to Earth) of the
  synthetic close approaching asteroids.  The dashed straight line
  is a linear fit to the data.}
\label{fig.close}
\end{figure}

Our study considers three different classes of asteroids that might be
identified soon after discovery as impactors:
\begin{itemize}

\item Impactors\\ 
We used a 133
member subset of the population of Earth impactors developed for
\citet{Veres2009} that strike the Earth in a 12 month period beginning
at the same time as the 12 month survey simulation described below
(\S\ref{ss.SurveySimulation}).  The impactors' orbit elements are
drawn from a realistic population of NEOs \citep{Bottke2002} and tend
to have perihelia or aphelia that lie near Earth's geocentric distance
of about $1\au$ (fig. ~\ref{fig.aei}).  These types of orbits are
nearly tangent to Earth's orbit so they spend more time available for
impact.  There is also an enhancement with small inclinations for the
same reason.

\item Close Approachers\\
We generated 8,275 synthetic asteroids from the \citet{Bottke2002} NEO
model that approach Earth to within 10\,LD (lunar distances,
$\la0.028\au$) during the one year simulated survey
(\S\ref{ss.SurveySimulation}). The orbit distribution of the close
approachers is more representative of the underlying NEO population
(fig.~\ref{fig.aei}) and has a higher mean eccentricity and
inclination compared to the impactors.  This leads to them having
higher speeds relative to Earth and higher apparent rates of motion as
viewed from ground-based observatories (compared to the impactors).
The number of close approachers increases linearly with the impact
parameter because the area of annuli of fixed width increases linearly
with the annuli diameter (fig.~\ref{fig.close}).

\item Main Belt Objects (MBO)\\ 
There are about one million asteroids larger than $1\km$ diameter in
the main belt with semi-major axes between about $2.0\au$ and $3.5\au$
(fig.~\ref{fig.aei}).  By definition, they can not approach within
$\sim0.7\au$ of Earth but tens of thousands will be within the
detection limits of the survey that we model below
(\S\ref{ss.SurveySimulation}) and many of them will not be known
objects (at least in the beginning of the survey).  We will show that
the objects' rates of motion are typical of some incoming impactors
and we wanted to determine if a relatively large astrometric
uncertainty could generate false non-zero impact probabilities even
for these distant objects.  We used a sample of about 14,000 synthetic
main belt asteroids selected from the \citet{Grav2011} solar system
model that have minimum perihelion magnitudes\footnote{The minimum
  perihelion magnitude is the apparent magnitude an object would have
  if observed at opposition from Earth when the object is at
  perihelion.} detectable in our synthetic survey ($V<20$). We used
the absolute magnitudes ($H$) from the \citet{Grav2011} model that
were assigned randomly according to a realistic size-frequency
distribution.

\end{itemize}

\subsection{Survey simulation}
\label{ss.SurveySimulation}

We simulated the detection of small incoming impacting asteroids with
the in-development ATLAS \citep{Tonry2011} system because of its
all-sky every-night survey capabilities.  The smaller the asteroid the
more likely it is that it will not be brighter than any of the
contemporary or planned survey system's limiting magnitudes until a
few days before impact, so that detecting the object requires nearly
all-sky coverage over that time interval.  ATLAS achieves all-sky
coverage by using small telescopes with wide fields-of-view (FOV) and
large-format CCD cameras.  Thus, it has a relatively large pixel scale
(and astrometric uncertainty) and brighter limiting magnitude than
other surveys.

The two primary purposes of this study were to 1) measure the
time-evolution of the impact probability of asteroids detected with a
realistic survey and 2) measure the effect of parallax on the impact
probability precision.

To address the first issue we generated synthetic observations of each
of our synthetic asteroids for a simulated one year ATLAS survey.  Our
low-fidelity instantiation of the survey covered the entire dark sky
each night without regard for the Moon, weather, galaxy and
clouds. The survey does account for the changing duration of the night
through the year and geometrical constraints from the horizon.  The
fidelity of the simulation is not critical to the two primary purposes
of this study but will have an impact on \eg\ the calculated detection
efficiency for small objects.

We addressed the second issue by using the synthetic survey to
simulate the performance of two ATLAS surveys located at observatory
sites F51 and 568 (respectively, the locations of the \PSone\ facility
on Haleakala, Maui, Hawaii, and the University of Hawaii 2.2\,m
telescope on Mauna Kea, Hawaii). These locations are amongst the best
ground-based astronomical sites in the world with typically $>75\%$
clear nights, dark sky, high altitude, sub-arc-second seeing, and a
number of other operational observatories and instruments. The sites
are separated by $\sim 130\km$ to enable parallax measurements for
nearby asteroids.  For instance, an asteroid at 10\,LD can have a
parallax of up to $\sim6.5\arcsec$ from the two sites --- about an
order of magnitude larger than the system's astrometric
uncertainty. Increasing the distance between the two observatories
would enhance the parallax effect. However, the meteorological
correlation would be lower and it would be less likely that both sites
could observe. Furthermore, the greater the separation between the two
sites the more difficult it is to survey the same fields.  Finally,
two separate sites introduces additional cost and management issues.

\begin{figure}[btp]
\centering

\epsscale{0.8}
\plotone{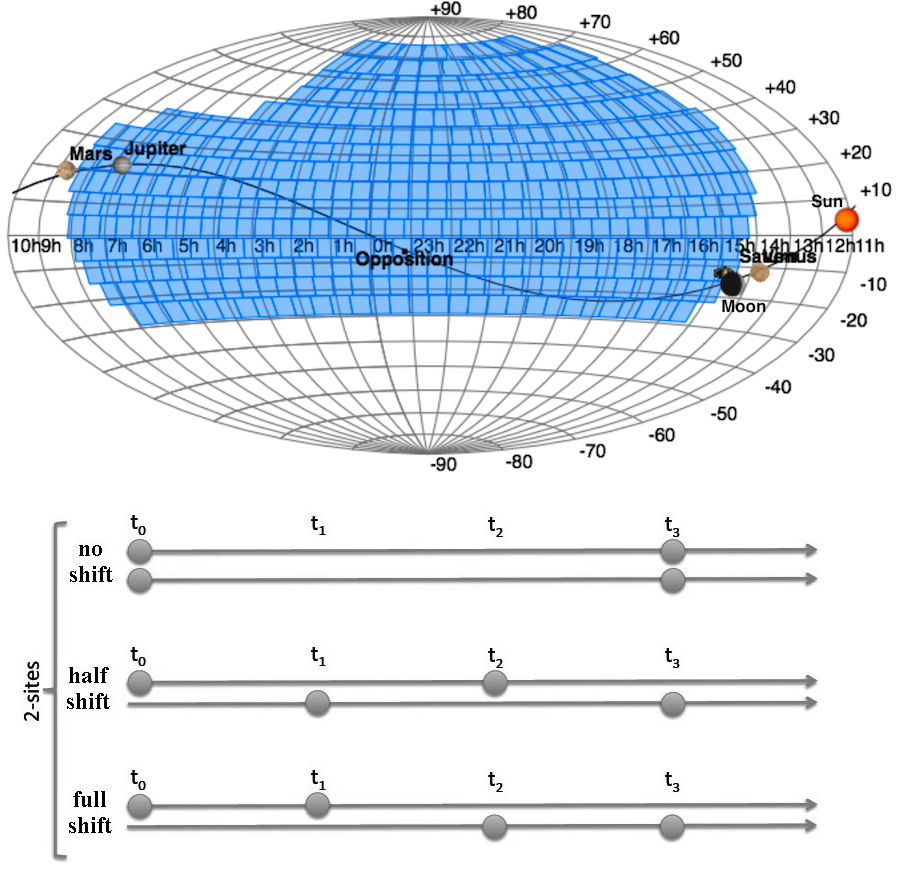}

\caption{(top) One night of the synthetic ATLAS survey covering the
  entire night sky visible from Haleakala, Maui, Hawaii.  Each shaded
  `square' represents one bore site that is imaged 4 times/night. The
  dark solid line represents the ecliptic and the positions of some of
  the planets, Sun and Moon are represented with their images.  The
  declination limit of $-30\arcdeg$ is roughly $40\arcdeg$ above the
  horizon as observed from Haleakala.  (bottom) Time series for
  detections in tracklets for the single site `quad' scenario and the
  2 site `no-shift', `half-shift' and `full-shift' scenarios.  Each of
  the time steps represent a transient time interval
  (TTI).}
\label{fig.survey}
\end{figure}

To create the survey we divided the sky into `square' tiles (or
fields) with each square having an area equal\footnote{Some of the
  values for the ATLAS system characteristics used in this work
  represent early expectations for the system.  The exact values have
  no impact on our general conclusions.} to that of an ATLAS camera's
FOV of $\sim40$\,deg$^2$. The tile centers were equally spaced at
about the width of the ATLAS FOV along lines of latitude.  They were
also spaced in latitude by the width of the ATLAS FOV.  This pattern
is not optimal because it results in significant field overlap at high
latitudes but it was simple to implement and the details of the survey
pattern will have little impact on the results.  We used the
\emph{Tools for Automated Observing optimization} (TAO)
package\footnote{Paolo Holvorcem, {\tt
    http://sites.mpc.com.br/holvorcem/tao/readme.html}} to schedule
the nightly surveying of the tiles and maximize the number of fields
exposed each night with the desired cadence subject to the survey's
limitations. Each tile was visited 4 times per night with roughly a
Transient Time Interval (TTI) of 15 minutes between visits.  We did
not account for the Moon, galactic plane, planets, bright stars or
weather (fig.~\ref{fig.survey}) but did account for the camera readout
and telescope slew times.  The fields were observed only when the Sun
was more than $12\arcdeg$ below the horizon and the field centers were more
than $30\arcdeg$ above the horizon (\ie\ above 2 airmasses).  We imposed a southern declination limit of $-30\arcdeg$ that will have the effect of decreasing the detection efficiency for the imminent impactors that are concentrated towards opposition, but will provide more time for surveying the `sweet spots' \citep{Chesley2004} at small solar elongations where the sky-plane density of future and larger impactors is highest.  The strategy of surveying the sweet spots is a likely scenario for actual surveys that are rightfully more concerned with the long-term advance notice of larger impactors rather than the short-term notice for smaller objects.

We used the \PSone\ Moving Object Processing System
\citep[MOPS,\ ][]{Denneau2013} to generate the synthetic asteroid
detections in our simulated survey and to link detections within a tile
on one night into `tracklets'.  MOPS employes an N-body integrator and
the DE406 ephemerides \citep{Standish1998} to compute the position and
brightness of every synthetic detection.  In practice, we set the
absolute magnitude of each synthetic object to $H=0$ and turned off
the MOPS system's capability of adding astrometric and photometric
uncertainty to each detection so that we could modify those values
{\it post hoc} as described below.

\subsection{Photometric \& astrometric uncertainty}
\label{ss.photometry_astrometry}

Each synthetic MOPS detection had a calculated apparent magnitude
$m_0$ corresponding to the value if the object's absolute magnitude
($H$) was zero.  We could then assign any other absolute magnitude to
the object and its {\it actual} apparent magnitude corresponding to
that detection would than be $m^*=m_0+H$.  We then used a {\it
  reported} apparent magnitude $m=m_0+H+\Delta m$ where $\Delta m =
{\tt max}\{0.01,G[0,\sigma(m^*)]\}$ and $G$ represents a randomly
generated number from a normal (Gaussian) distribution with a mean of
zero and width $\sigma(m^*)=0.02\times2^{(m^*-16)}$ appropriate for
the ATLAS system. The {\tt max} function limits the minimum
photometric uncertainty to 0.01\,mag.

\begin{figure}[btp]
\centering
\epsscale{0.4}
\plotone{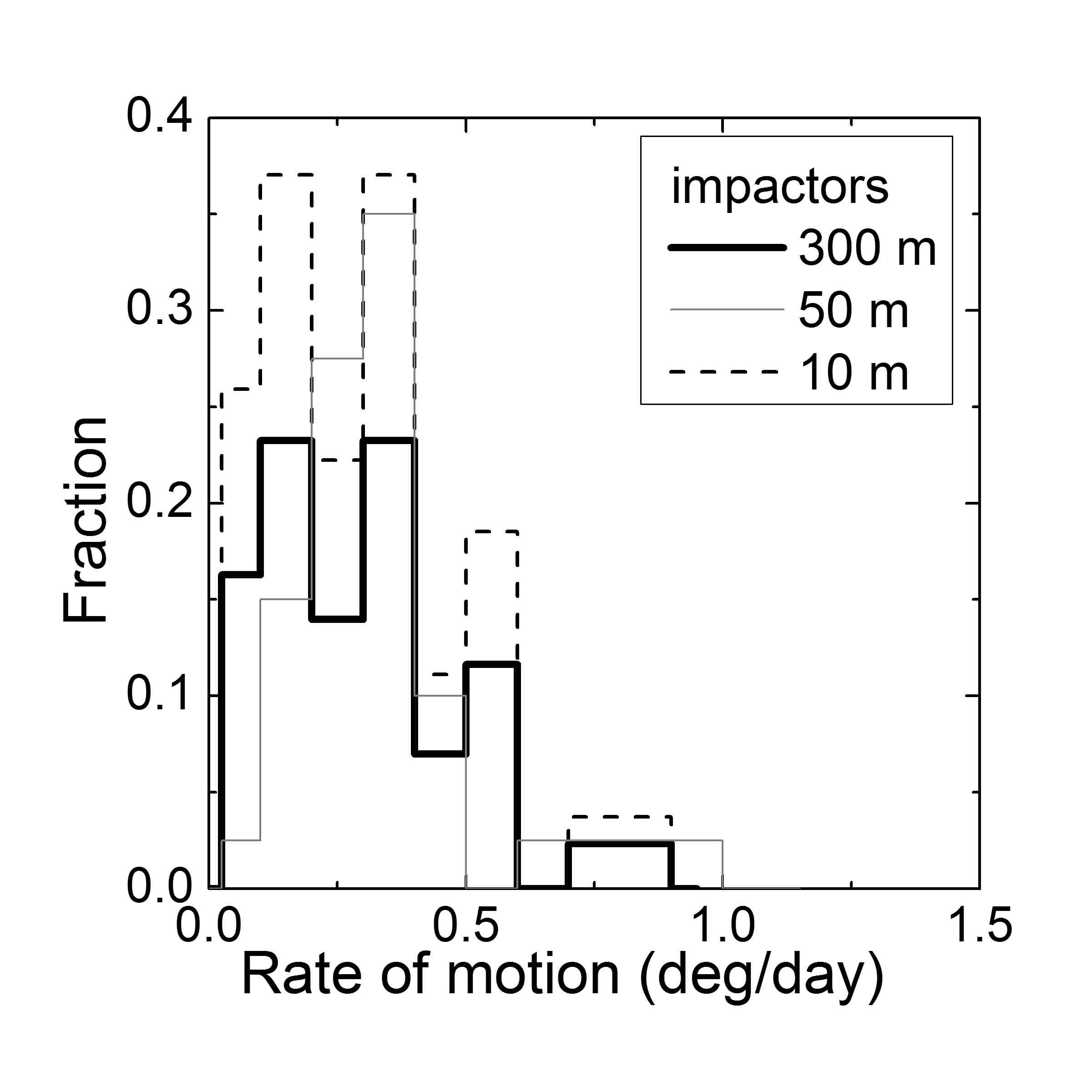}
\epsscale{0.4}
\plotone{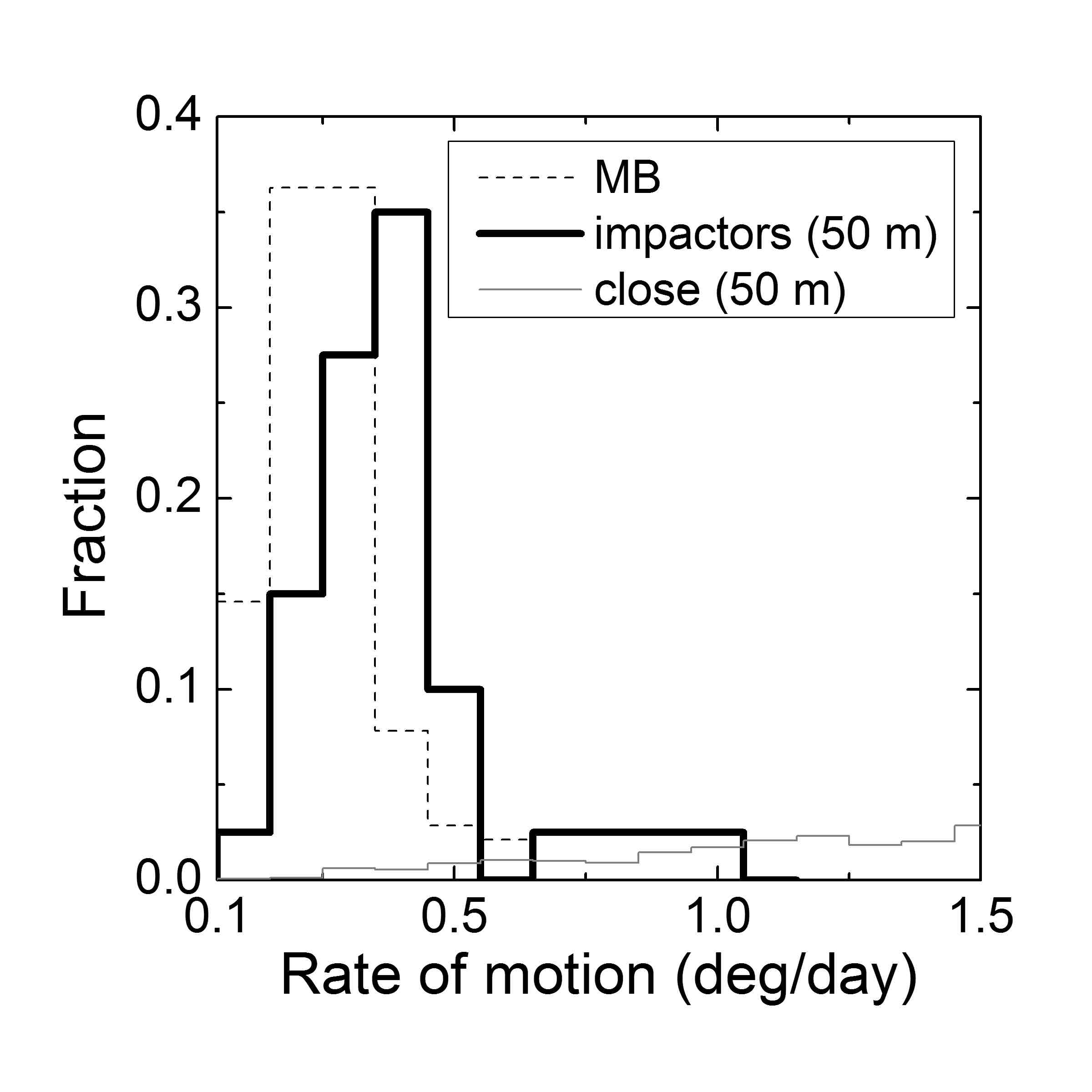}

\caption{(left) Apparent rate of motion on the first night of
  observation for impactors with diameters of $10\meter$, $50\meter$
  and $300\meter$.  (right) Apparent rate of motion on the first night
  of observation for impacting, close approaching and main belt
  asteroids.  The impactors and close approachers both have diameters
  of $50\meter$ but the main belt asteroids have a realistic
  size-frequency distribution.}
\label{fig.rates}
\end{figure}

We did not account for the effect of trailing of the detections due to
the motion of the object during an exposure
\citep[\eg\ ][]{Veres2012}.  The neglect is justified because the
large ATLAS plate scale of about 2$\arcsec$/pixel means that
detections trail by $<1$~pixel for rates of up to 1.6$\arcdeg$/day and
the majority of the impactors that are $>10\meter$ diameter move
slower than this rate when they are first detected
(fig.~\ref{fig.rates}).  Main belt objects are not trailed at all
because they typically move fastest near opposition at rates of about
0.25$\arcdeg$/day.  The close approachers can move much faster but
here we assume that ATLAS will apply a trail finding and fitting
algorithm to maintain astrometric and photometric integrity for
objects with faster rates motion.

The impact probability calculation depends on the reported astrometric
uncertainty that, in turn, depends on the apparent brightness of the
detections and the system's pixel scale.  The reported astrometric
position of each synthetic detection was `fuzzed' by an offset
$\Delta={\tt max}\{0.1\arcsec,G[0,\sigma_p(m^*)]\}$ where the
sub-script `p' represents `positional' uncertainty and
$\sigma_p(m^*)=2\arcsec/10^{8.5-0.4m^*}$ appropriate to the ATLAS
system's $2\arcsec$ pixel scale and photometric performance.  The {\tt
  max} function ensures that the minimum astrometric uncertainty is
always $>0.1\arcsec$.

\subsection{Survey cadence}
\label{ss.SurveyCadence}

One of the main goals of this study was to demonstrate how two survey
sites could improve the impact probability determination in comparison
with a single site, but there are many different survey cadences that
could be implemented in either scenario.  We decided that a fair
comparison between the two scenarios required maintaining the same
combined tracklet arc-length (\ie\ time from first to last
observation) and studied three different 2-site visit cadences (see
fig.~\ref{fig.survey}):

\begin{itemize}

\item 1-site (quads) \\
4 images acquired with roughly a TTI between each

\item 2-site no-shift \\
2 images acquired at each site with roughly 3 TTI between them

\item 2-site half-shift \\ 
2 images acquired at each site with roughly 2 TTI between them 
interleaved with the other site

\item 2-site full-shift \\
2 images acquired at each site with roughly a TTI between them 
sequential with the other site

\end{itemize}

We will show in \S\ref{ss.IPevolutionForImpactors} that there is little difference in performance between the cadences but there is a marginal benefit to the 2-site full-shift scenario and we adopt it as the nominal cadence unless otherwise specified.

\subsection{Observability windows}
\label{ss:ObservabilityWindow}

For a fair comparison between the the observation circumstances of our
synthetic close approachers and impactors we only considered
observations of close approachers {\it before} the moment of closest
approach.  This requirement is symmetric with the impactors because it
is impossible to obtain observations of an impactor after impact.

Furthermore, because our simulated survey was only one year in
duration instead of infinite, we were careful to consider only those
impactors of different sizes that could be detected before impact.  To
do so we defined an impactor diameter ($D$) dependent `time
observability window', $t_{window}(D)$, and then, letting $t_{begin}$
and $t_{end}$ represent the simulated survey's starting and finishing
time respectively, require that the time of impact of an object with
diameter $D$, $t_{impact}(D)$, satisfy $t_{impact}(D) \ge t_{begin} +
t_{window}(D)$ and $ t_{impact}(D) \le t_{end}$.  \ie\ we require that
the time of impact {\it and} the entire observability window be in the
simulated survey time.

\begin{figure}[btp]
\centering
\epsscale{0.8}
\plotone{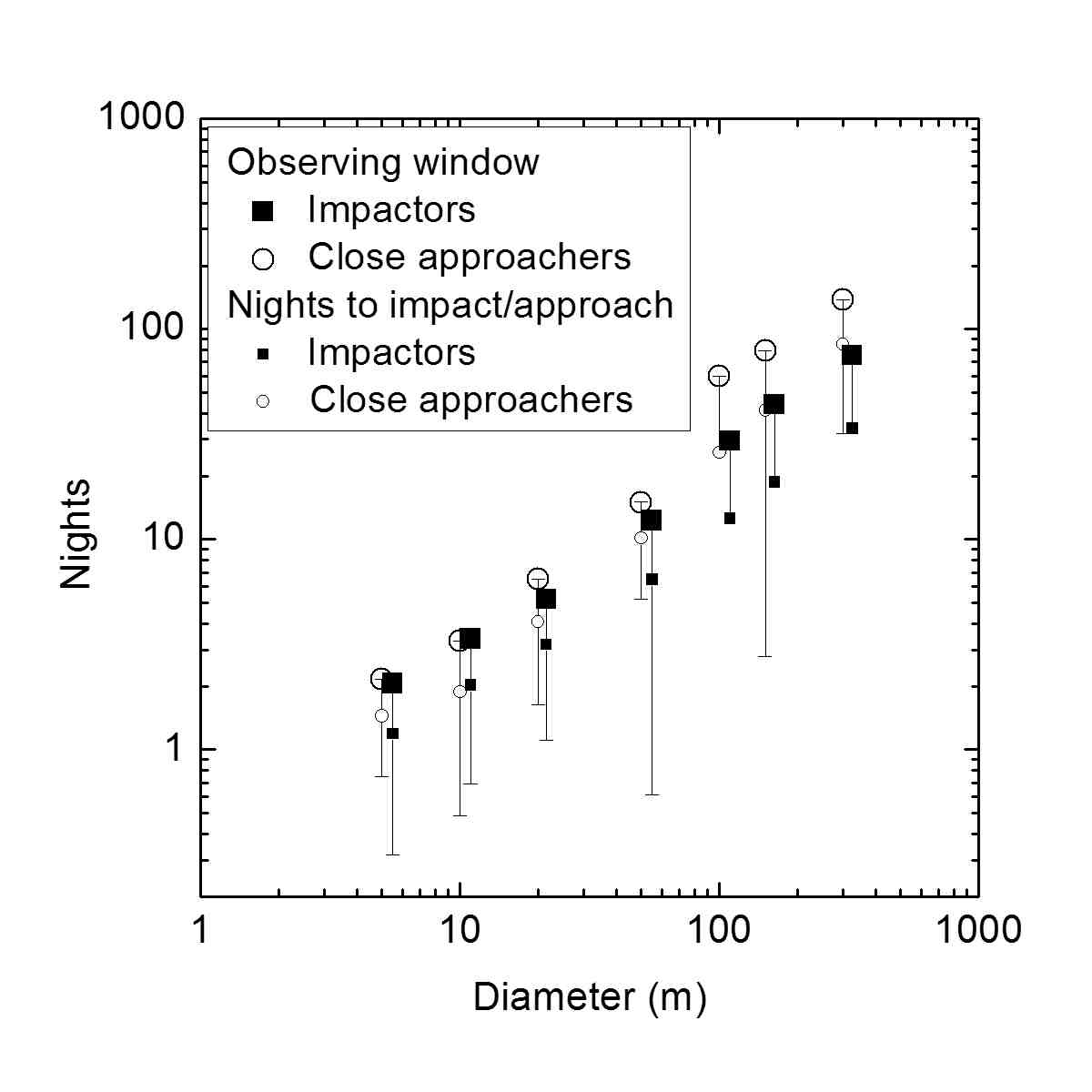}
\caption{(large circles and squares) Observing window duration implemented in this study for close approachers and impactors on their final approach.  (small circles and squares) Average number of nights to closest approach or impact.  The upper limits of the error bars on these data points corresponds to the observing window durations by definition (\S\ref{ss:ObservabilityWindow}).}
\label{fig.time_windows}
\end{figure}

We defined $t_{window}(D)$ using our simulated survey and synthetic
objects from which we could measure the number of days between the
first detection and impact as a function of the impactor diameter
(fig.~\ref{fig.time_windows}).  If $\bar t_{first}(D)$ represents the
average value as a function of diameter and $\sigma_{first}(D)$ the
standard deviation of the distribution then we set $t_{window}(D) =
\bar t_{first}(D) + \sigma_{first}(D)$.  This time observability
window encompasses the actual observability window of about 84\% of
the objects at each diameter but will eliminate the $\sim 16$\% of the
sample with the longest observability times.  The choice reflects a
balance between increasing the time observability windows and keeping
more objects in the analysis --- longer windows mean fewer objects
satisfy the requirement.  No window was applied for main belt
asteroids.

\subsection{Orbit determination and impact probability}
\label{s:od}

Orbit determination is the process of identifying the best-fit
least-squares orbit to the astrometric dataset. We used the standard
differential correction procedure \citep[Chap.~5]{milani_orbdet} with
the objects's synthetically generated orbit as the starting point.
The use of the synthetic orbit as the initial orbit in the fit will
skew our orbit determination and impact probability calculation
towards more accurate values than could be expected from the
single-survey performance in our study.  We do not consider this an
important issue because the NEO candidate followup community rapidly
provides additional astrometry for initial orbit determination by the
Minor Planet Center.  It is also worth noting that we assume a high
efficiency and accuracy for linking detections of the same object into
tracklets and then linking the tracklets into `tracks'
\citep[Chap.~8]{milani_orbdet} but this is justified because the MOPS
has $>$99.5\% efficiency at doing so \citep{Denneau2013}.

After an orbit was computed along with its corresponding uncertainty
we computed the probability of an Earth impact. The analysis of close
encounters is typically a strongly non-linear problem whose solution
requires sophisticated methods \citep[\eg,][]{Milani2005} but we
adopted a simplified approach to the calculation of the impact
probability since this study only assesses its evolution in the days
and weeks before impact.  First we search for upcoming close
approaches and then we perform a linear mapping of the orbital
uncertainty region to the close encounter $b$-plane and compute the
impact probability corresponding to the intersection between the
mapped uncertainty region and Earth's cross section
\citep{Valsecchi2003}. The short-term propagation and the lack of
intervening planetary encounters justifies the adoption of the
simplified linear approach.

We updated the orbit determination and risk assessment night-by-night
within the simulation to assess the time evolution of the impact
probability. \ie\ we incrementally added each tracklet to an object's
astrometric data set and then recalculated the orbit and associated
impact probability each night.

\section{Results \& Discussion}

We tested the performance of the ATLAS survey for impactors and close
approachers of 5, 10, 20, 50, 100, 150, and 300$\meter$eters diameter
because we expected there to be a size-dependency on the ability to
rapidly calculate impact probabilities.  Large objects will typically
be detected at larger distances where the effects of parallax are
small and the lever arm to impact is large, making it difficult to
assign a high probability to a possible impact.  Small objects will be
detected close to Earth so parallax will provide some power in the
orbit solution and impact probability calculation.

\subsection{Observable characteristics of the impactor population}

The mean
apparent rate of motion of impactors on the first night of detection
is about $0.25\pm0.05$~deg/day, essentially matching both the mean
rate and distribution of typical main belt asteroids moving at
$0.19\pm0.06$~deg/day (fig.~\ref{fig.rates}).  The fastest impactor rates of $\ga
0.5$~deg/day mimic that of the perfectly harmless Hungaria asteroids
on the inner edge of the main belt
\citep[\eg][]{Jedicke1996,Rabinowitz1991} that can also have very high
rates of motion in ecliptic latitude because of their high
inclinations.  On the other hand, the Earth-grazing close approachers
have high apparent fly-by speeds even on their night of first
detection ranging up to $10$~deg/day and $<2$\% have rates of
$<0.5$~deg/day in the range of the typical impactor.

\begin{figure}[btp]
\centering

\epsscale{1.0}

\plotone{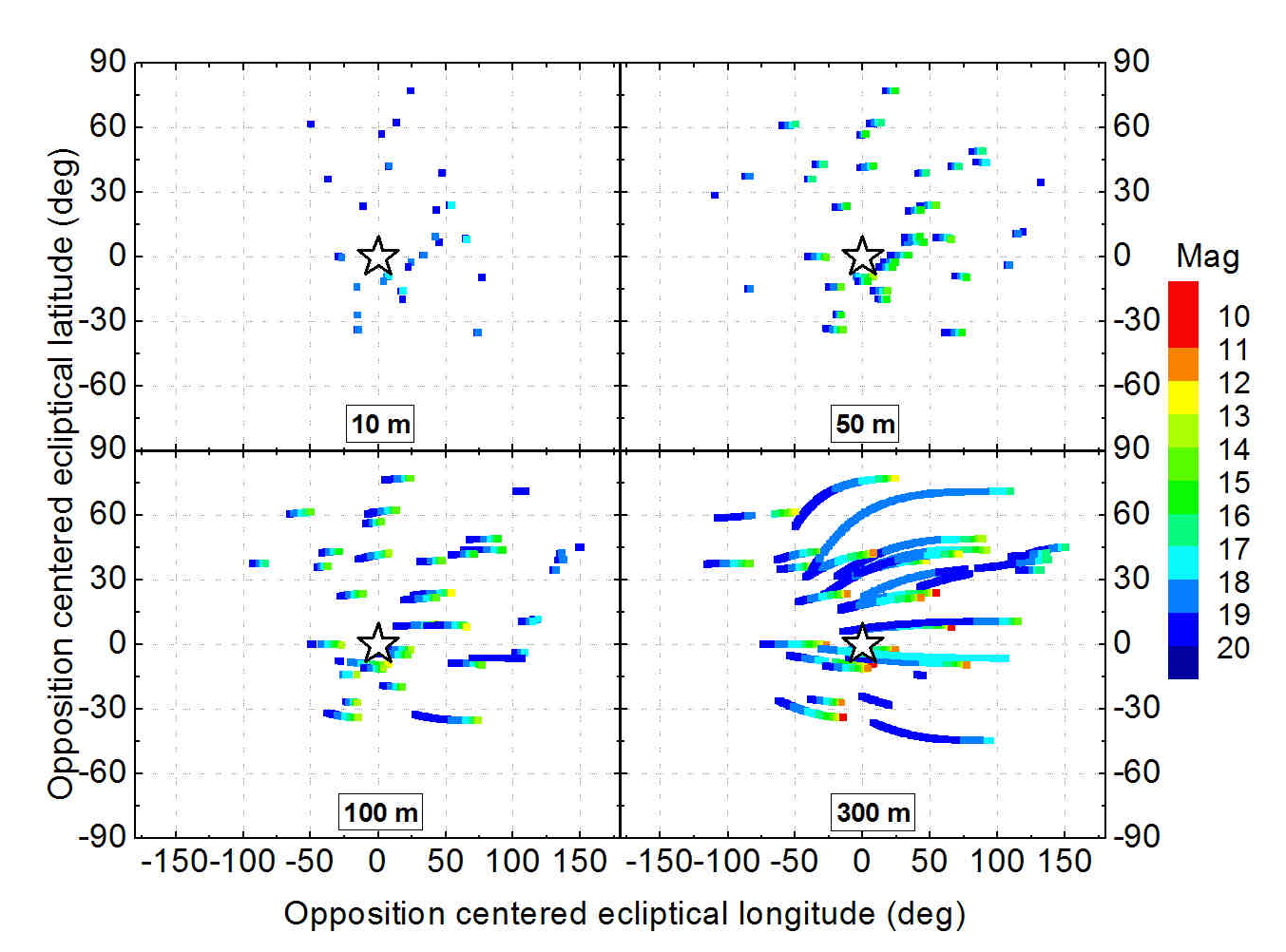}

\caption{Time evolution of the sky plane position and apparent $V$
  magnitude for all the detected synthetic impactors at 4 different
  sizes: $10\meter$, $50\meter$, $100\meter$ and $300\meter$.  The
  coordinates are ecliptic opposition-centric with west to the
  right. The star symbol in the center represents
  opposition.}
\label{fig.impactors-skyplane-motion}
\end{figure}

Figure~\ref{fig.impactors-skyplane-motion} illustrates several
features of the detected impactors's sky plane motion.  First, smaller
objects are visible for much less time than the larger impactors, the
impactors typically move westwards and mostly in longitude, and they
are first detected near the system's limiting magnitude and increase
in brightness as they move closer to Earth. The larger objects can be
discovered almost everywhere on the night sky but the smaller objects
tend to be discovered towards opposition where the reduction in
apparent brightness due to phase angle effects is smallest.

\subsection{Detection efficiency \& rates}

\begin{figure}[btp]
\centering
\plottwo{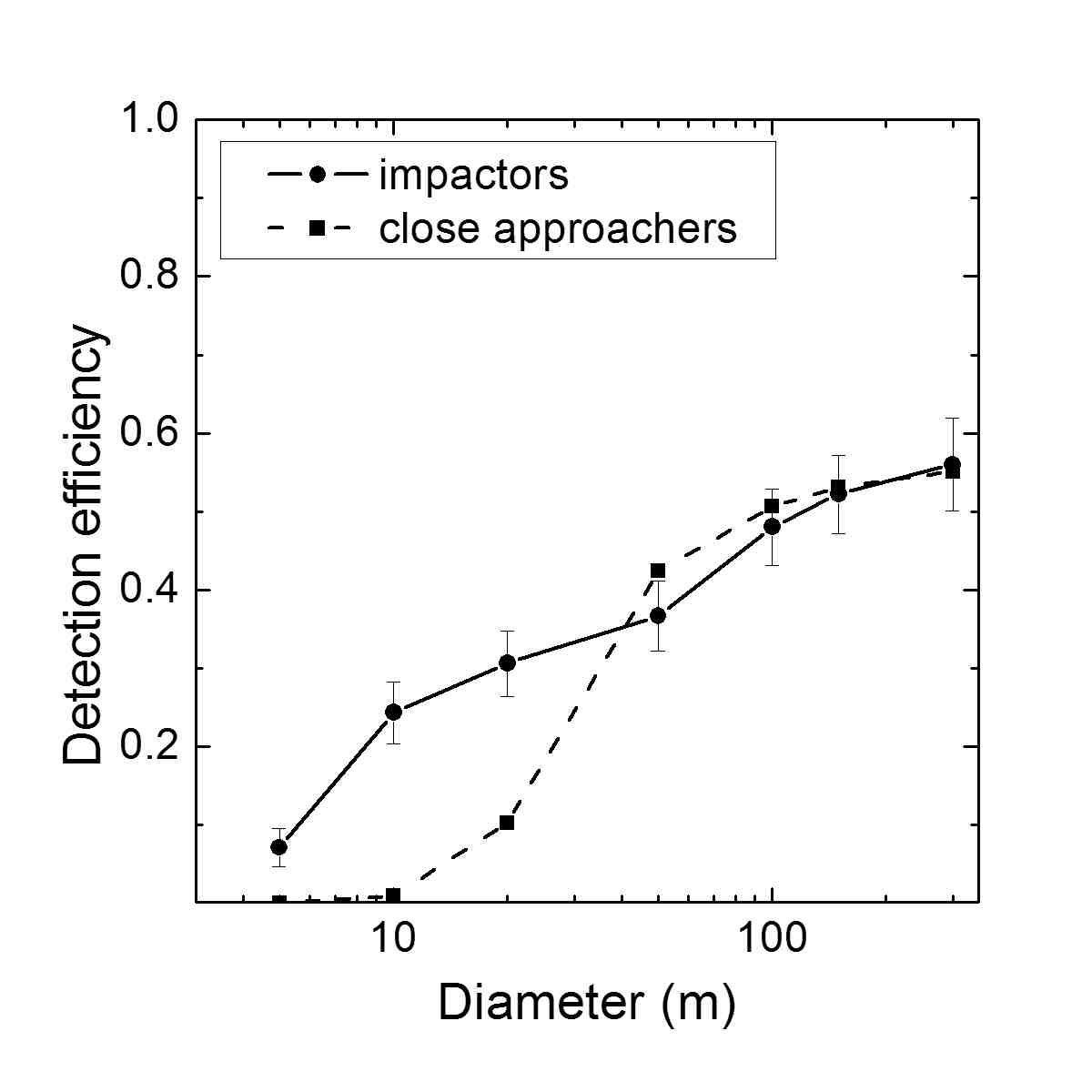}{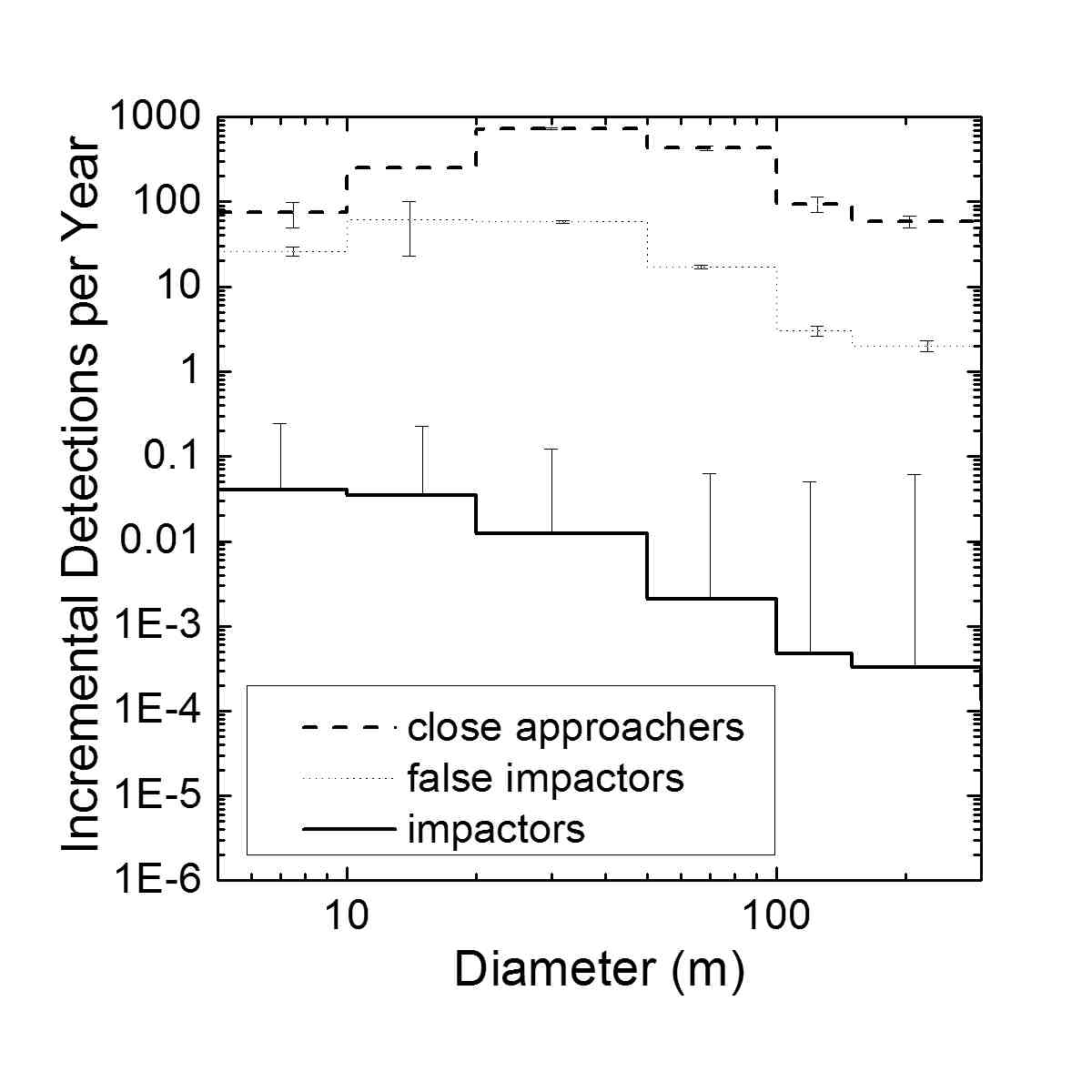}

\caption{ (left) Detection efficiency for synthetic impactors and close
  approachers as a function of object diameter and (right) our
  predicted incremental number of detections per year of impactors,
  close approachers, and false impactors (close approachers with a residual impact probability of $>10^{-6}$ on the day of closest approach --- see fig.~\ref{fig.close-approachers}). Our calculation of the discovery rates used the \citet{Brown2013}
  impactor size-frequency distribution and impact rate.  Note that the close approacher statistics only includes objects {\it before} closest approach for direct comparison to the impactors.}
\label{fig.eff}
\end{figure}

The ATLAS pre-impact detection efficiency (figs.~\ref{fig.eff})
plateaus at a maximum of about 50\% even for the largest objects for
geometrical reasons --- ATLAS only surveys about half the sky but
detects everything brighter than its limiting magnitude.  The
efficiency decreases for smaller objects with a particularly dramatic
drop from $10\meter$ to $5\meter$.  \ie\ it decreases by about half,
to about 25\%, from $300\meter$ to $10\meter$ diameter --- nearly 3
orders of magnitude in the impactors cross-sectional area --- but then
decreases by $>50$\% from $10\meter$ to $5\meter$ diameter (only a
factor of 4 in cross-section). In this size range the objects are
typically well below the survey system's limiting magnitude and even a
nightly cadence is not sufficient to catch the objects in the brief
time they are bright enough to be detected before impact.

The behavior of the detection efficiency as a function of diameter is
different for close approaching asteroids (figs.~\ref{fig.eff}).
Remembering that for a direct comparison to the impactors we only
considered close approachers detected {\it before} closest approach,
the detection efficiencies are identical at the largest sizes we
considered.  The close approacher detection efficiency would have been
much higher had we allowed them to be detected after close approach
because there would be much more time to detect them and because
objects that approach in daylight from the direction of the Sun are
likely to depart and be detectable in the night time sky.  The
detection efficiency is slightly higher for the close approachers in
the $50\meter$ to $150\meter$ diameter range because these objects
will be bright enough and detectable longer than the impactors.  On
the other hand, close approachers of $<50\meter$ diameter are detected
less efficiently than the impactors because they are usually too far
away and therefore too faint to be detected --- 1\% of the close
approachers in our simulation come no closer than $1$\,LD while all
the impactors do so by definition.

Figure~\ref{fig.eff} illustrates that it is unlikely that ATLAS will
detect impactors larger than 5\,m diameter.  There are simply not
enough of them and the detection efficiency for the smaller, most
frequent, impactors is $<10$\%.  It is dangerous to extrapolate
ATLAS's ability to detect even smaller impactors in the 1-2\,m
diameter size range like \TC\ \citep[\eg][]{Jenniskens2009} because
the detection efficiency becomes particularly sensitive to the
observing cadence and subtleties of image processing \eg\ trail
detection.  However, assuming that the detection efficiency drops to
just 1\% for a $1\meter$ diameter object ATLAS might detect one
impactor every few years.

On the other hand, ATLAS will detect 100s or 1,000s of close
approaching asteroids to within 10\,LD because there are far more of
them than impactors (scaling like the ratio of the cross sectional
area of a circle with radius equal to the close approach distance and
the cross sectional area of Earth).  Most of the detected close
approachers will be in the $20\meter$ to $50\meter$ diameter range
(fig.~\ref{fig.eff}) but there will still be 10s to 100s of detected
objects of $<20\meter$ and $>50\meter$ diameter.  We will show in
\S\ref{ss.IPforMBandCAs} that ATLAS alone can not establish that all
the close approachers are not impactors --- 100s of the closest
approaching objects will always have a residual non-zero impact
probability unless additional observations are acquired with other
optical or radar facilities.

\subsection{Impact probability evolution for impacting asteroids}
\label{ss.IPevolutionForImpactors}

\begin{figure}[btp]
\centering
\plotone{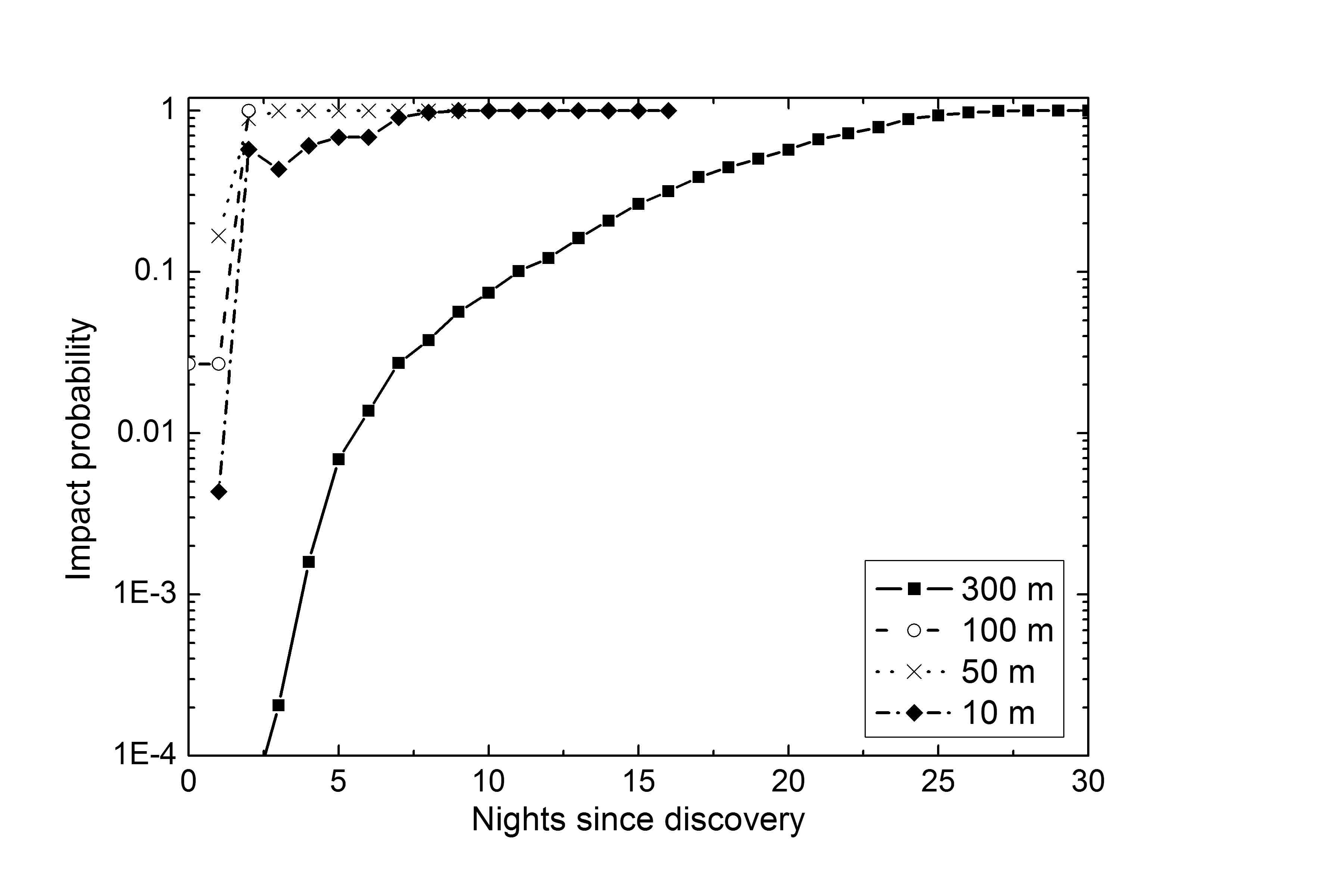}
\caption{Impact probability time evolution for four synthetic objects of
  $10\meter$, $50\meter$, $100\meter$ and $300\meter$
  diameter.}
\label{fig.impactProbabilityEvolution}
\end{figure}

The impact probability for an impacting asteroid depends on the
observed arc length and the object's distance from Earth and both
depend on the size of the impactor (see
fig.~\ref{fig.impactProbabilityEvolution}). Longer arc lengths provide
better orbit precision and a higher impact probability because all
these objects are actually impactors. However, when an object is far
from Earth the integration to the impact epoch stretches the orbital
uncertainty and decreases the impact probability.  The impact
probability is smaller for larger diameter impactors at the same
number of nights after discovery because smaller objects are likely to
be observed when close to Earth and therefore closer to their impact
time.  The topocentric parallax in the detections and deviation from
great-circle motion allow a better orbit determination and the small
amount of time to impact provides a higher impact probability.

The typically monotonically increasing impact probability that
asymptotically approaches unity
(fig.~\ref{fig.impactProbabilityEvolution}) is the expected behavior
though we were surprised that the single-system astrometry required
about a week to reach $\sim100$\% for the $50\meter$ diameter
impactors and a month to reach the same values for the $100\meter$
diameter impactors.  The impact probability reached 99\% an average of
$2.0\pm0.5$, $4.5\pm1.8$, $8.9\pm4.7$ and $28.7\pm16.3$ days before
impact for impactors of $10\meter$, $50\meter$, $100\meter$, and
$300\meter$ diameter at mean geocentric distances of $0.6\pm0.4\,LD$,
$1.0\pm0.5\,LD$, $1.9\pm1.0\,LD$ and $5.5\pm1.5\,LD$ respectively.
Thus, even though a $300\meter$ diameter object has nearly
$1000\times$ the cross-sectional area of a $10\meter$ diameter object,
and is therefore about a $1000\times$ brighter ($\sim7.5\mags$) at the
same topocentric distance, an imminent impact becomes definitive with
only about $10\times$ more warning time when the larger object is only
$\sim5\times$ further away.

Of course, our calculated impact probabilities rely only on the
detections from a single survey with relatively large astrometric
uncertainty compared to contemporary standards (because of its all-sky
coverage as described in \S\ref{ss.photometry_astrometry}).  In some
cases there will be nights with particularly bad astrometric error
that temporarily decrease the impact probability as illustrated by the
100$\meter$ diameter example in
fig.~\ref{fig.impactProbabilityEvolution}.  The detection of a real
impactor with even a $10^{-6}$ impact probability would certainly
trigger high-precision ground-based optical and radar followup that
would improve the orbit determination and impact probability.
  
None of the four objects represented in
fig.~\ref{fig.impactProbabilityEvolution} had an impact probability
calculated on the first night.  This does not mean that the impact
probability was zero, only that with just one night's data the orbit
determination could not converge (even with our method where we {\it
  start} with the correct initial orbit) on a full 6-parameter orbit
along with its covariance matrix and both are required for the impact
probability calculation.  While short-arc orbit determination methods
are available they have not yet been extensively tested in their ability to provide reliable hazard
assessments \citep[\eg][]{Virtanen2001, Milani2004, Chesley2005}.
With our single survey simulation there is a $\sim14$\% efficiency for
calculating an impact probability on the first night for 10$\meter$
diameter objects with the efficiency decreasing with diameter to just
2\% for 300$\meter$ impactors. Therefore, it is critical to have rapid
follow-up observations from other observatories to achieve a better
efficiency in recognizing impactors on the discovery night --- but
there may be nothing particularly noteworthy about the tracklet to
flag it as worthy of immediate follow-up.

\begin{figure}[btp]
\centering

\plotone{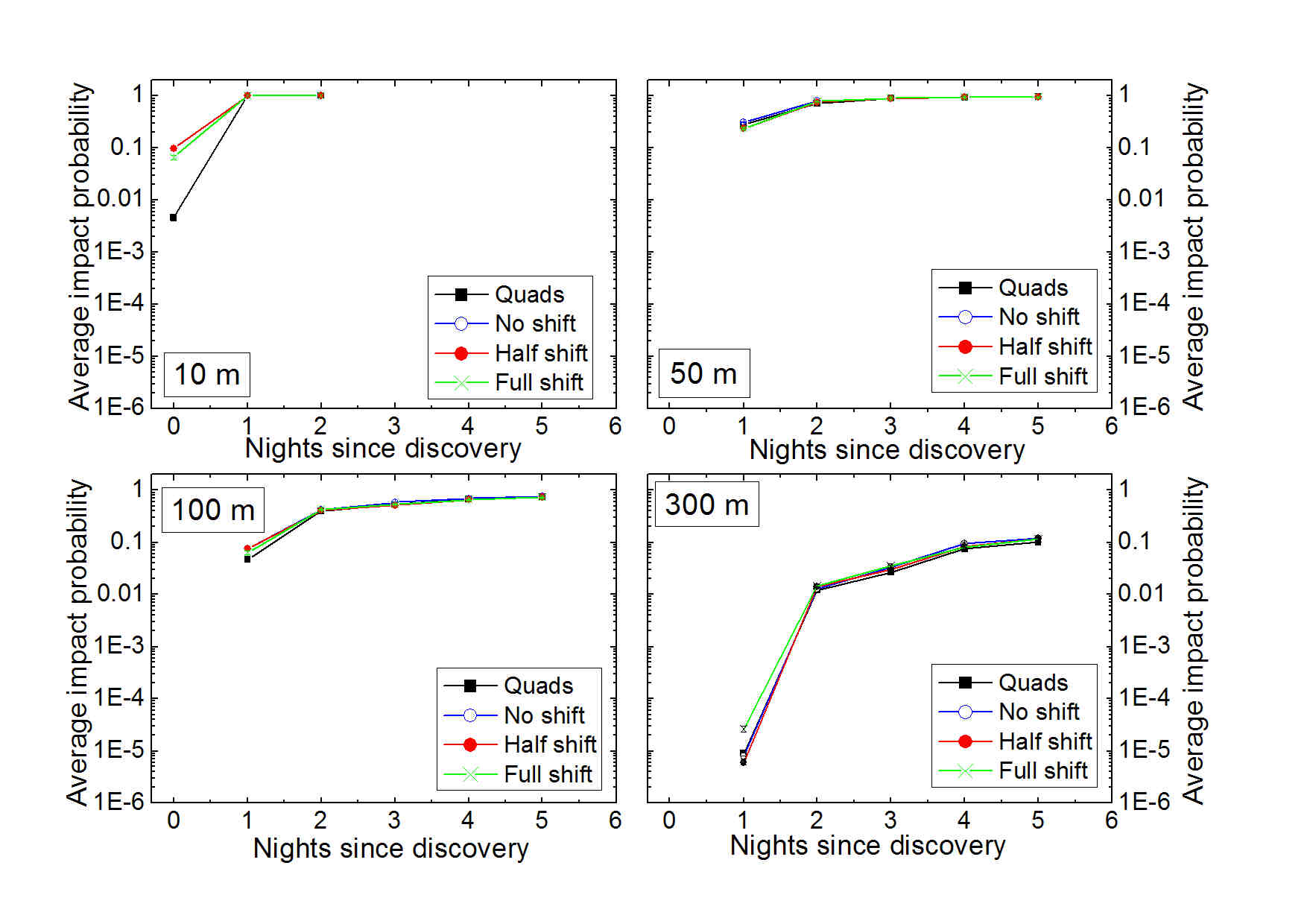}

\caption{Average
  impact probability as a function of the number of nights since discovery
  for impactors of $10\meter$, $50\meter$, $100\meter$
  and $300\meter$ diameter. The four curves correspond to the cadence
  scenarios illustrated in fig.~\ref{fig.survey} and discussed in
  \S\ref{ss.SurveyCadence}. The missing data points for night zero indicates that it was not possible to calculate the impact
  probability on the discovery night: this is not the same as claiming the objects have zero
  impact probability. }
\label{fig.averageIP.vs.night}
\end{figure}

The average impact probability as a function of time for an ensemble of impactors of a
specific diameter (fig.~\ref{fig.averageIP.vs.night}) behaves much the
same as the individual examples illustrated in
fig.~\ref{fig.impactProbabilityEvolution}.  The four survey cadences
(\S\ref{ss.SurveyCadence}) are essentially equivalent in terms of the
numerical value and efficiency of the impact probability calculation
beginning on the second night after discovery.  While the efficiency
for calculating an impact probability on the first night is small
(discussed above), the 2-site full-shift scenario is always superior,
so that combining observations from different stations does provide
better constraints on the orbit and allows for an impact probability
calculation. Again, we stress that this is true only on the first
night of discovery and only for the small fraction of objects for
which an impact probability can be calculated, so the benefits of
parallax in the impact probability calculation afforded by a 2-site
scenario is limited to a small fraction of the least dangerous
impactors.  We recognize that the 2-site scenario offers other
advantages such as increased immunity to weather shutdowns (one site
can still survey if the other is in-operational) and natural disasters
(such as lava flows) but at the expense of having to maintain two
sites.

\begin{figure}[btp]
\centering
\epsscale{0.8}
\plotone{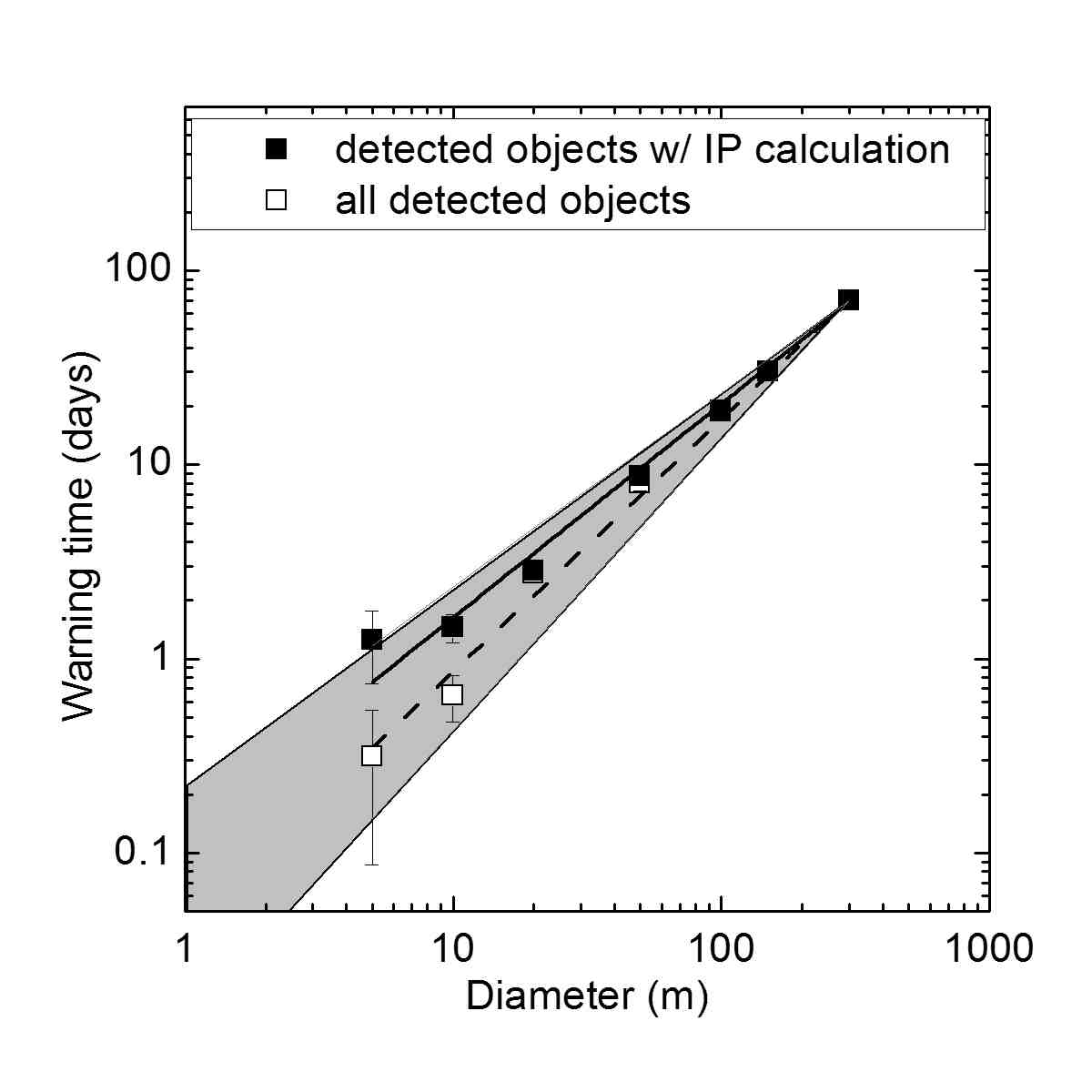}
\caption{Impact warning time $t_{warn}$ as a function of impactor diameter ($D$) for the
ATLAS survey using the full-shift cadence (see fig.~\ref{fig.survey}
and \S\ref{ss.SurveyCadence}).  The error bars represent the standard
error on the mean and are equal to or smaller than the data points for all but the two leftmost values.  (dashed line) The fit to the
data for all detected impactors (\ie\ including those without calculated orbits and impact probabilities) with warning time given by 
$\log_{10}(t/\mathrm{days}) = (1.3 \pm 0.1)\,\log_{10}(D/\mathrm{meters}) - (1.4 \pm 0.2)$. 
(solid line) The fit to the objects with calculated impact probability is 
$\log_{10}(t/\mathrm{days}) = (1.1 \pm 0.2)\,\log_{10}(D/\mathrm{meters}) - (1.0 \pm 0.4)$. 
The grey area represents the expected range with slopes in the range $[1.0,1.5]$ (see \S\ref{ss.IPevolutionForImpactors}) when anchored at 
$300\meter$ diameter.}
\label{fig.close_approach_warning_time}
\end{figure}

We define the impact warning time as the time interval between the
impact epoch and the epoch at which an object's Palermo Scale
\citep{Chesley2002} ranking becomes $>-2$. We used the Palermo Scale
because it is a standard tool to communicate the risk posed by a
possible impact and selected the $-2$ threshold as it corresponds
to cases that `merit careful
monitoring'.\footnote{http://neo.jpl.nasa.gov/risk/doc/palermo.html}.   The warning time increases with the impactor diameter as $t_{warn}
\propto D^{1.1 \pm 0.2}$ (fig.~\ref{fig.close_approach_warning_time})
--- {\it detected} impactors {\it with} an orbit determination that are smaller than $\sim20\meter$
diameter typically have less than a couple of days of warning time,
50$\meter$ diameter objects provide about one week notice, and the
warning time is weeks to months for objects $>100\meter$ diameter.  Including those objects that were detected but for which an orbit was impossible to calculate (even starting with the correct orbit as the initial value in the fitting procedure) the warning time increases with 
diameter as $t_{warn} \propto D^{1.1 \pm 0.2}$.

The time to impact from first detection ($\Delta t$) should increase
linearly with diameter $D$ because the geocentric distance
($\rho_{limit}$) at which an asteroid becomes brighter than a system's
limiting magnitude ($m_{limit}$) is given by
\begin{equation}
5\log_{10} \rho_{limit} = m_{limit} - H + \phi(r,\rho_{limit})
\end{equation}
where $\phi$ is a `phase function' that depends on the objects
geocentric and heliocentric ($r$) distance \citep{Bowell1988}.
Furthermore, $H \propto \log_{10} D$ \citep{Pravec2007} so it follows
that $\rho_{limit} \propto D$ assuming that $\phi$ is roughly constant
(which is justified given that the heliocentric distance is nearly
constant during the final approach ($r \sim 1$) and because $\phi$
depends on the phase angle, which does not change much during an
impacting object's final approach to Earth).  Since $\rho_{limit} = v
\Delta t$ where $v$ is the speed of the impactor relative to Earth it
follows that $\Delta t \propto D$.

However, $v$ depends on the diameter because smaller objects are
detected closer to Earth where they are moving faster because they
have accelerated in Earth's gravity well.  Assuming that impactors
fall towards Earth with similar initial $v_\infty$ then it is not
difficult to show that the speed at discovery goes roughly as $v
\propto D^{1/2}$ and that $\Delta t \propto D^{3/2}$.

The impact warning time $t_{warn}$ as we have defined it is related,
but not identical to, and always $\le \Delta t$.  Thus, we expect that
$t_{warn} \propto D^x$ with $1<x<1.5$ in agreement with our measured
value of $1.3\pm0.1$.  To put this in perspective, a Chelyabinsk-like
impactor of $20\meter$ diameter would typically only have a 2\,day
warning time {\it if} it was detected by an ATLAS-like survey.

\subsection{Impact probability evolution for main belt 
and close approaching asteroids}
\label{ss.IPforMBandCAs}

We included a sample of main belt and close approaching asteroids to
assess the survey's ability to distinguish them from impactors.  The
problem may be difficult when only a short arc of observations is
available in which case the orbital uncertainty is large and there may
be multiple and very different initial orbits \citep{Milani2008}
consistent the observations, some of which lead to a least-square
solution far from the actual orbit.

The inclusion of the distant, slow-moving main belt asteroids may seem
surprising but fig.~\ref{fig.rates} illustrates that the rate of
motion of impacting asteroids can be similar to that of main belt
objects.  Despite the similarity in their rates of motion we never
found a non-zero impact probability for a synthetic main belt object
with $\ge2$\,days of arc.  It is worth noting once again that we used
the actual main belt orbit as the starting point to obtain the orbit
solution --- this process biases our results towards decreasing the
calculated impact probability a short time after discovery.

The most likely false impactors must be PHOs that experience a close
Earth approach.  These objects might be identified as they approach
Earth and the astrometric uncertainties and orbit integration may
combine to produce non-zero impact probabilities.  Indeed, about 30\%
of the close approaching objects in our synthetic population had an
impact probability $> 10^{-6}$ at some time during the
simulation\footnote{$10^{-6}$ is the threshold typically used by NASA
  to rule out an impact. \eg\ {\tt
    http://www.jpl.nasa.gov/asteroidwatch/newsfeatures.cfm?release=2013-017}}
but none of the impact probabilities ever exceeded 3\%.

\begin{figure}[btp]
\centering
\epsscale{1.0}

\plotone{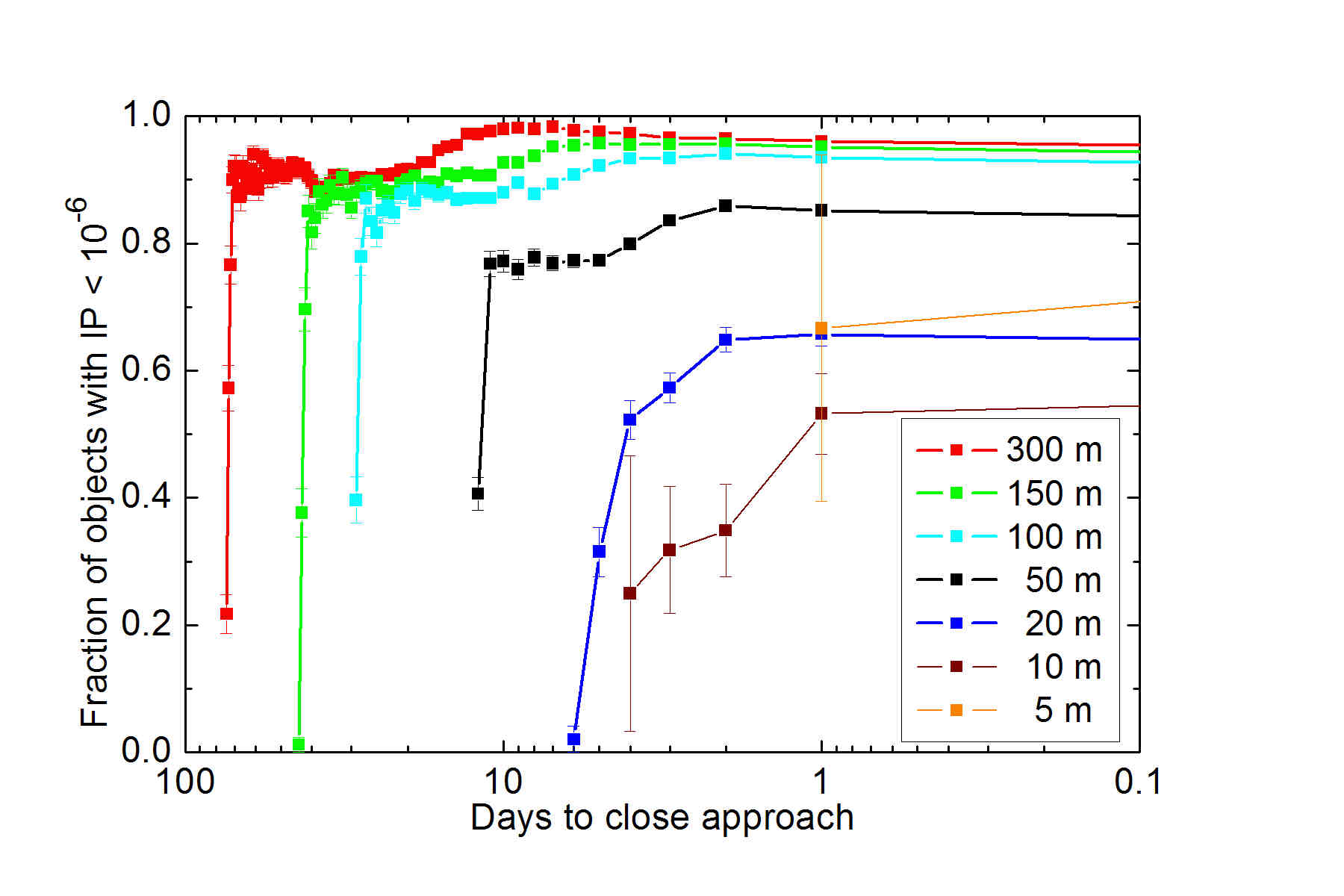}
\caption{Fraction of close approaching asteroids for which an impact
  is ruled out (impact probability $< 10^{-6}$) as a function of the
  number of days before close approach for four different asteroid diameters.  The  `noisy' behavior on
the left, corresponding to long times before impact, is mostly due to low number
statistics.}
\label{fig.close-approachers}
\end{figure}

With our single-system survey simulation (\ie\ one or two sites) we
find that even large objects, those $\ge50\meter$ diameter, can have
non-zero impact probabilities just a few days before impact
(fig.~\ref{fig.close-approachers}).  About 3 to 5\% of the $50\meter$
and $100\meter$ diameter close approachers have a persistent impact
risk {\it on} the day of (false) impact which means that follow-up
observations from other stations are critical to establish the lack of
danger from these objects.  A persistent impact risk remains on the
day of impact for about 75\% of the $10\meter$ diameter close
approachers but perhaps this is not too worrisome since they are
unlikely to make it through Earth's atmosphere and cause serious
ground damage.  On the other hand, the opportunity for scientific
study of more \TC-like events \citep{Jenniskens2009} is tremendous if
the false alarm rate is small enough so, once again, follow-up
observations are required for all these objects.

We estimated the false impactor rate (fig.\ref{fig.eff}, right panel)
with the single-survey system using the \citep{Brown2013} PHO size
frequency distribution appropriate for objects in this size range, our
calculated close approacher survey efficiency (fig.~\ref{fig.eff},
left panel), and the fraction of them that retain a non-zero impact
probability on the (false) impact date
(fig.~\ref{fig.close-approachers}).  We find that the single all-sky
survey system will generate 100s of false impactors per year for
objects of $\la100\meter$ diameter.  Thus, rapid astrometric followup
with other optical and radar facilities is imperative to reduce the
false impactor rate to zero.

\section{Conclusions}

We have performed a simulation of a single all-sky asteroid survey to
study the time evolution of the calculated impact probability for both
real and false impactors.  We also studied the utility of using two
observatories at different locations to perform the survey to take
advantage of the parallactic displacement in the detections of the
same object.

As expected, the impact probability for impactors typically increases
monotonically with time after discovery and is larger at the time of
discovery for small objects that are detected closer to Earth and with
less time to impact.  We found that the impact warning time, the time
interval between when the impact probability reaches -2 on the Palermo
Scale and when the impact takes place, increases with diameter
according to $t_{warn} \propto D^{1.3}$ and developed a simple
mathematical argument that the exponent should be in the range
$[1.0,1.5]$.

Close approaching asteroids can almost always be unambiguously
identified as non-impactors but a small percentage will have a
non-zero impact probability even on the day of (false) impact.
\ie\ the simulated survey on its own is unable to eliminate the impact
risk.  The fraction of objects for which a persistent impact risk
exists at the time of impact increases with decreasing diameter of the
object because the small objects have smaller observational arc
lengths and concomitantly less precise orbit elements.  The
combination of the PHO size-frequency distribution with the
probability of detecting false impactors suggests that the single
all-sky system alone will generate 100s of potential impactors that
must be ruled out with other followup facilities.

The calculated impact probability can take surprisingly long to reach
$\sim100$\% with just the results from a single low-precision
astrometric survey.  The impact probability may reach 100\% only a few days
before impact even for $300\meter$ diameter objects detected a
month in advance and imaged nightly thereafter .

Our simulations suggest that a 2-site survey is unnecessary, at least
in terms of the incremental benefit in improving the impact
probability calculation.  The parallax afforded by this scenario only
improves the impact probability calculation for a small fraction of
the smallest asteroids detected shortly before impact.  The 2-site
survey offers many different cadence options and some can provide more
efficient impact probability calculations than others.  The derived
impact probability was $\sim10\times$ higher (\ie\ better) on the
discovery night using the `full-shift' cadence compared to the other
two cadences.  This suggests that a real survey that implements the
2-site scenario should carefully test different cadences to select one
that maximizes the efficiency and accuracy of the impact probability on
the discovery night.  The effect of survey cadence on the impact
probability calculation is negligible on successive nights.

\acknowledgements

\noindent{\bf Acknowledgements}

We thank J. Tonry and L. Denneau of the ATLAS survey for assistance in
characterizing and designing a representative ATLAS survey.
D. Farnocchia was supported for this research by an appointment to the
NASA Postdoctoral Program at the Jet Propulsion Laboratory, California
Institute of Technology, administered by Oak Ridge Associated
Universities through a contract with NASA. Peter Vere\v{s}'s
Pan-STARRS MOPS Postdoctoral Fellowship was sponsored by the NASA NEOO
grant No. NNX12AR65G.

\bibliographystyle{pasp}
\bibliography{references}




\clearpage

\clearpage

\clearpage

\clearpage

\clearpage

\clearpage

\clearpage

\clearpage

\clearpage

\clearpage

\clearpage

\end{document}